\documentclass[conference]{IEEEtran}

\usepackage{amsmath,amssymb,amsfonts}
\usepackage[linesnumbered,ruled,vlined]{algorithm2e}
\usepackage{graphicx}
\usepackage{textcomp}
\usepackage{listings}
\usepackage{fancyhdr}
\usepackage{comment}
\usepackage{mathptmx} % This is Times font
\usepackage{soul}
\usepackage{multirow}
\usepackage{booktabs}
\usepackage[normalem]{ulem}
\usepackage[sort,nocompress]{cite}
\usepackage[T1]{fontenc}
\usepackage{amssymb}% http://ctan.org/pkg/amssymb
\usepackage{pifont}% http://ctan.org/pkg/pifont
\usepackage{tablefootnote}
\usepackage[T1]{fontenc}

\usepackage[hyphens]{url}
\usepackage[final]{microtype}
\usepackage[table,xcdraw,dvipsnames]{xcolor}
\usepackage[keeplastbox]{flushend}

\usepackage{enumitem}
\setlist[enumerate]{noitemsep}

\usepackage[bookmarks=true,breaklinks=true,letterpaper=true,colorlinks,linkcolor=blue,citecolor=blue,urlcolor=blue]{hyperref}
\newcommand{\insertFigure}[2]{
    \begin{figure}[t!]
        \centering
        \includegraphics[width=\linewidth]{figures/#1.pdf}
	\vspace{-6mm}
        \caption{\small #2}
	\vspace{-3mm}
        \label{fig:#1}
    \end{figure}
}

\newcommand{\insertWideFigure}[2]{

    \begin{figure*}[ht!]
        \centering
        \includegraphics[width=\textwidth]{figures/#1.pdf}
	\vspace{-6mm}
        \caption{\small #2}
	\vspace{-3mm}
        \label{fig:#1}
    \end{figure*}

}

\newcommand{\squishlist}{
 \begin{list}{$\bullet$}
  { \setlength{\itemsep}{0pt}
     \setlength{\parsep}{0pt}
     \setlength{\topsep}{3pt}
     \setlength{\partopsep}{0pt}
     \setlength{\leftmargin}{1.5em}
     \setlength{\labelwidth}{1em}
     \setlength{\labelsep}{0.5em} } }
     
\newcommand{\squishnums}{
 \begin{list}{$\bullets$}
  { \setlength{\itemsep}{0pt}
     \setlength{\parsep}{3pt}
     \setlength{\topsep}{3pt}
     \setlength{\partopsep}{0pt}
     \setlength{\leftmargin}{1.5em}
     \setlength{\labelwidth}{1em}
     \setlength{\labelsep}{0.5em} } }

\newcommand{\squishlisttwo}{
 \begin{list}{$\bullet$}
  { \setlength{\itemsep}{0pt}
     \setlength{\parsep}{0pt}
    \setlength{\topsep}{0pt}
    \setlength{\partopsep}{0pt}
    \setlength{\leftmargin}{2em}
    \setlength{\labelwidth}{1.5em}
    \setlength{\labelsep}{0.5em} } }

\newcommand{\squishend}{
  \end{list}  }

\newcommand{\greencheck}{{\color{ForestGreen}\checkmark}}
\newcommand{\orangecheck}{{\color{orange}\checkmark}}
\newcommand{\redcheck}{{\color{red}\xmark}}

\newcommand{\DataflowName}[0]{\textsc{PipeOrgan}~}
\newcommand{\DataflowNameTitle}[0]{PipeOrgan}

\newcommand{\DataflowNamenospace}[0]{\textsc{PipeOrgan}}

\newcommand{\NocName}[0]{\textsc{AMP}~}

\newcommand{\NocNamenospace}[0]{\textsc{AMP}}

\newcommand{\TODO}[1]{\textcolor{red}{TODO::: #1}}
\newcommand{\YC}[1]{{\color{purple}\bfseries [Yu-Hsin::: #1]}}
\newcommand{\RG}[1]{{\color{orange}\bfseries [RG::: #1]}}
\newcommand{\TK}[1]{{\color{olive}\bfseries [Tushar::: #1]}} 
\newcommand{\HK}[1]{{\color{teal}\bfseries [Jun::: #1]}}
\newcommand{\LL}[1]{{\color{brown}\bfseries [Liangzhen::: #1]}}
\newcommand{\EQ}[1]{{\color{cyan}\bfseries [Eric::: #1]}}

\newcommand{\xmark}{\ding{55}}%

\newcommand{\insertWideFigureScaled}[3]{

    \begin{figure*}[ht!]
        \centering
        \includegraphics[width=#3\textwidth]{figures/#1.pdf}
	\vspace{-6mm}
        \caption{\small #2}
	\vspace{0mm}
        \label{fig:#1}
    \end{figure*}

}
\usepackage{url}
\usepackage{authblk}

\title{\textsc{\DataflowNameTitle}: Efficient Inter-operation \underline{Pipe}lining with Flexible Spatial \underline{Organ}ization and Interconnects
}

\makeatletter
\renewcommand\AB@affilsepx{, \protect\Affilfont}
\makeatother

\author[1]{Raveesh Garg}
\author[2]{Hyoukjun Kwon}
\author[3]{Eric Qin}
\author[3]{Yu-Hsin Chen}
\author[1]{Tushar Krishna}
\author[4]{Liangzhen Lai}
\affil[1]{Georgia Tech}
\affil[2]{UC Irvine}
\affil[3]{Meta}
\affil[4]{Witmem}

\begin{document}
\date{}

\maketitle
\thispagestyle{plain}   %%%%%
\pagestyle{plain}
\begin{abstract}

Because of the recent trends in Deep Neural Networks (DNN) models being memory-bound (e.g., large language models and convolutional neural networks with heavy skip connections), inter-operator pipelining for DNN accelerators is emerging as a promising optimization. Inter-operator pipelining reduces costly on-chip global memory and off-chip memory accesses by forwarding the output of a layer as the input of the next layer within the compute array, which is proven to be an effective optimization by previous works.

However, the design space of inter-operator pipelining is huge, and the space is not yet fully explored. In particular, identifying the right depth and granularity of pipelining (or no pipelining at all) is significantly dependent on the layer shapes and data volumes of weights and activations, and these are different even within a domain. For instance, AR/VR applications have 6 orders of magnitude swing in the activation to weight ratios. Another factor affecting the right depth and granularity is the dependencies in the form of skip connections, which increase the activation accesses and vary in reuse distance and density across applications.

Moreover, works divide the substrate into large chunks and map one layer onto each chunk, which requires communicating halfway through or through the global buffer. However, for fine-grained inter-operation pipelining, placing the corresponding consumer of the next layer tile close to the producer tile of the current layer is a better way to exploit fine-grained spatial reuse.

In order to support variable number of layers (ie the right depth) and support multiple spatial organizations of layers (in accordance with the pipelining granularity) on the substrate, we propose \DataflowNamenospace, a new class of spatial data organization strategy for energy efficient and congestion-free communication between the PEs for various pipeline depth and granularity. \DataflowName takes advantage of flexible spatial organization and can allocate layers to PEs based on the granularity of pipelining. We also propose changes to the conventional mesh topology to improve the performance of coarse-grained allocation.
\DataflowName achieves 1.95x performance improvement over the state-of-the-art pipelined dataflow on XR-bench workloads.
%\RG{Placeholder}

\end{abstract}

\vspace{-2mm}
\section{Introduction}
\label{sec:introduction}

Deep Neural Networks (DNN) are gaining popularity due to their use in applications including natural language processing (NLP)~\cite{devlin2018bert,achiam2023gpt}, computer vision~\cite{hrvit,resnet,liu2021swin} and personalized recommendations~\cite{naumov2019deep}. The most compute-intensive Deep Neural Networks operators are the Einstein summation (einsum)-based operators, which refers to dot product-based operations generalized to arbitrary dimensions (e.g., general matrix multiplications, or GEMM). Example DNN operators based on Einstein summation include linear layer, convolution, and batched matrix multiplication. Because the einsum-based operators typically account for 70\% of total latency on GPUs~\cite{sigma}, many accelerators focusing on matrix multiplication~\cite{tpu-isca,eyeriss2016isca,kwon2018maeri,extensor,sigma} and design-space exploration tools~\cite{kwon2021heterogeneous,kwon2019understanding,interstellar,timeloop,kao2020gamma,cosa} emerged.

DNN accelerators employ various dataflows and scheduling strategies to map the matrix multiplication spatially over the processing elements (PEs) and temporally. However, optimizing individual matrix multiplication operators does not always translate to optimal execution of the whole application. This is because it misses out on the opportunity to reuse the output feature map in the next operator that certain layers have. 
Therefore, current works are actively investigating inter-layer pipelining or inter-operator (operator means tensor-operator) pipelining in order to reuse the portions of the intermediate feature maps~\cite{tangram,fusedcnn,yan2020hygcn,simba,isca-pip,genetic-pipeline,tileflow}. Moreover, recent prior works also explore design-space exploration with inter-operator pipelining~\cite{atomic-dataflow,genetic-pipeline,tileflow,isca-pip}. Prior work FLAT~\cite{flat} and TANGRAM~\cite{tangram}, respectively claim 1.5x and 2x performance gains from pipelining, over their respective state-of-the-arts. However, the benefits of pipelining heavily depend on two aspects: (1) pipelining depth (i.e., how many layers do we pipeline) and (2) pipelining granularity (i.e., the tile size of the pipelining).

\insertWideFigureScaled{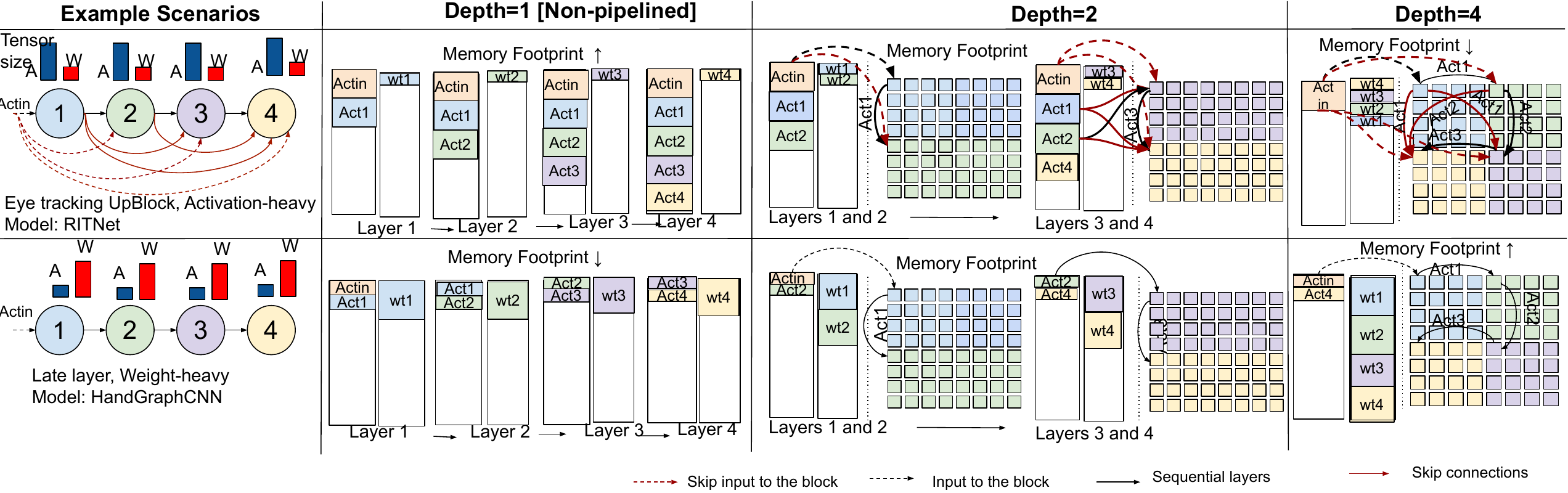}{Impact of pipeline depth for different sets of layers. We only show data movement with one PE from each layer in the figure. The boxes represent memory footprint, and not the space allocated in the buffer. Tensors with larger volumes get reused within PE array with deeper pipelining.\vspace{-4mm}}{1}

%\insertWideFigureScaled{motivation-2}{Role of spatial organization in fine-grained pipelining.
%}{0.8}

\insertWideFigure{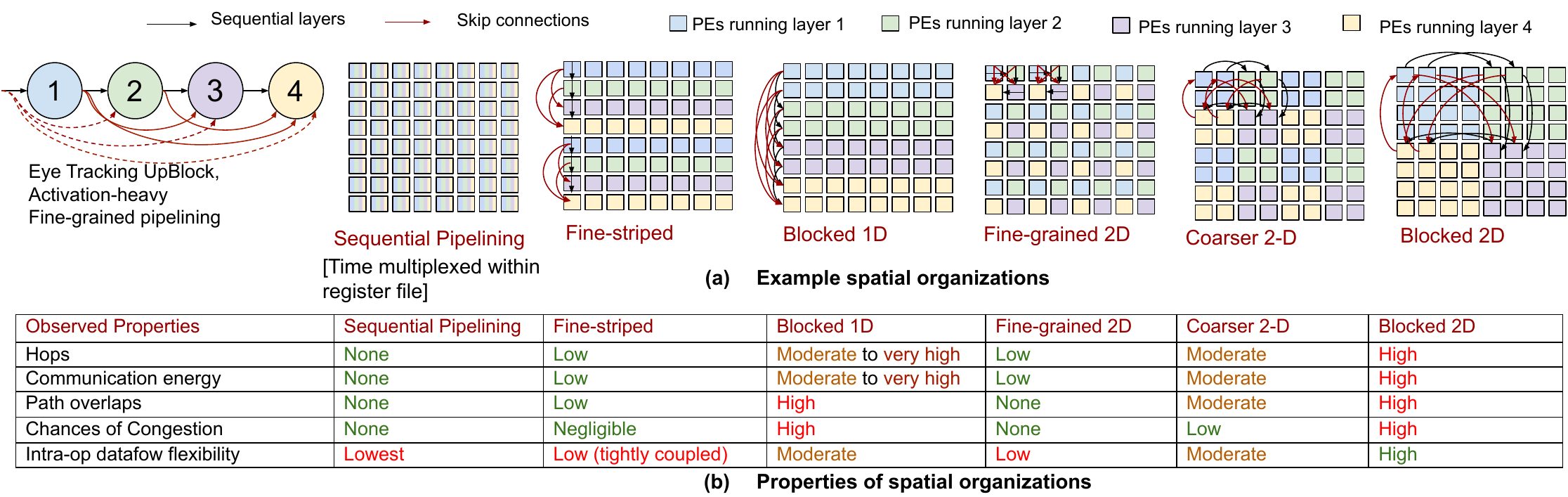}{High-level example showing the impact of spatial organizations on energy and latency. We consider depth = 4, along with the traffic that results after running an activation-heavy UpBlock in RITNet~\cite{eyeseg}, used extensively in eye-tracking. We only show data movement with 2 PEs from each layer. Cycle-level analysis on traffic is shown in~\autoref{fig:fine_grained_patterns}-\ref{fig:2Dinterleaving}\vspace{-3mm}}

%\insertFigure{organ-motivation}{Benefits of Spatial Organization based on depth and granularity.}

\noindent
\textbf{Pipelining Depth}%\footnote{Refers to number of layers being pipelined at a time}:}
~\autoref{fig:motivation-1} shows the impact of depth on two opposite kinds of layers. The right pipeline depth depends on two factors - A/W ratio ($Activation\,volume/Weight\,volume$) and skip connections.

\textit{A/W ratio:}Activation-heavy layers prefer higher pipelining depth, since, pipelining essentially reuses activations, which are shared between two consecutive layers, while shallow pipelining results in a large overhead of re-fetching big activations. On the other hand, weight-heavy layers do not favor pipelining, and prefer intra-operator reuse, since weights are not shared between the layers, and deeper pipelines, involves those unshared weights being together at a time. In case of models~\cite{handtracking,eyecod,fbnet,eyeseg,midas,keyword,emformer,planercnn,fasterrcnn,fbnetv3,tcn,hrvit,depthrefinement,planercnn} inside XR-bench~\cite{xrbench}, $Activation \, volume/Weight \, volume$ ratio roughly spans from $10^{-3}$ to $10^3$ as~\autoref{fig:AW} shows in~\autoref{sec:dg}.

\textit{Skip connections:} Skip connections make the otherwise activation-heavy layer, even more activation-heavy, since these combine activations from multiple preceeding layers. The last layer of the DenseNet~\cite{densenet} (used in RITNet~\cite{eyeseg} model for eye segmentation) block combines four activations, and op-by-op operator in the middle means, re-fetching all these activations.  Multiple XR-bench models have skip connections with different densities and reuse distances as shown in~\autoref{fig:skip} in~\autoref{sec:dg}. Skip connections also diminish the benefits if the pipeline is not deep enough. As~\autoref{fig:motivation-1} shows, with pipeline depth of 2, the output of layer 1 eventually needs to be written back because of a downstream dependency, and we only observe the benefit on output of layer 3.

\noindent
\textbf{Pipelining Granularity%\HK{The texts in the footnote need to go into the main text. Also, the definition in the footnote is not very clear; need to give a clearer definition.} 
%\footnote{Refers to coarseness and fineness of staging the layers}
:} Pipeline granularity refers to the size/portion of the intermediate tensor consumed by the consumer. In prior works on inter-operator pipelining, one layer is mapped on a block of PEs with each partition running an individual layer. This is inefficient for highly fine-grained pipelining between layers, and this spatial organization can lead to higher hop energy and possibly NoC congestion. ~\autoref{fig:orgs} Blocked-2D and Fine-grained-2D show examples of aforementioned spatial arrangements.
Interleaving creates a trade-off between locality and flexibility since finest-grained interleaving makes the layers tightly coupled to each other and poses constraints on the tile sizes. On the other hand, coarse grained interleaving, may not exploit enough locality but it provides, flexibility in tiling layers. We show that arrangement of the layers spatially depends on the preferred dataflow of the layers which also depends on the layer shape.

%~\autoref{fig:organ-motivation} shows few examples of spatial organization strategies with different depths and granularities of pipelining, while ~\autoref{fig:orgs} shows spatial organization strategies for depth=4.

In this work, we propose~\DataflowNamenospace\footnote{\underline{Pipe}lining \underline{Organ}ization}, a new class of spatial organization strategies for inter-operator pipelining, and a systematic optimization methodology for that. \autoref{fig:orgs} shows some examples of different organization strategies for depth=4.

We also characterize the possible traffic patterns that arise from plethora of factors, like depth, granularity, skip connections, load balancing layers with unequal MACs etc. Mesh topology does not achieve the best energy efficiency different kinds of traffic patterns, specially ones which involve skip connections. Moreover it requires us to trade-off on-chip energy for intra-operator flexibility, in cases where fine-grained pipelining is not possible, since coarse-grained allocation requires large hops. Also, spatial organizations on mesh topology trade-off inter-operator and intra-operator reuse specially in cases of large pipelining depths. Flattened butterfly~\cite{flattened} topology allows direct communication between distant PEs, however is an overkill, and increases the link complexity to $O(NlogN)$. Thus we propose changes to mesh topology that reduce congestion and hop count for mesh without adding too many or too long links.

Our contributions are as follows:
\squishlist
\item We show the importance of considering variable depth and pipelining, given variation across applications and variation within an application itself. 

\item We quantitatively show that activation/weight ratio is the key metric that affects the pipeline depth and granularity.

\item We propose~\DataflowNamenospace, a new class of spatial organization strategies, with different granularities at which multiple layers are arranged, ranging from fine-grained checkerboard and striped to traditional blocked 1-D or 2-D arrangement as~\autoref{fig:orgs} shows. This is the first work (to the best of our knowledge) that proposes finer-grained spatial organization strategies between different layers.

\item We characterize the traffic patterns resulting from different spatial organization strategies and different factors like depth, granularity, skip connections, unequal PE allocation between layers etc. We identify the bottlenecks in mesh and propose~\NocNamenospace\footnote{\underline{A}ugmented \underline{M}esh for \underline{P}ipelining}, a modified mesh for supporting coarse-grained patterns as well.
%\item We also propose a new NoC~\NocName\footnote{NoC Expansion}, an area efficient NoC that accelerates variable depth and granularity of pipelining for a myriad of traffic patterns.

%\item We also propose~\DataflowNameenospace,  a schedling methodology that determines the depth and granularity of pipelining various groups of layers within a model.

%\item \DataflowName and~\NocName provides xx reduction in energy and xx speedup over the state-of-the-art.%\RG{Placeholder}
\squishend
%Depth, Granularity

%Fig 1a Figure regarding

%FIg 1b GB

%Fig 1c NoC congestion

%Fig 1d Proposed design

\section{Background and Related Work}
\subsection{Individual Operation Dataflows and Mappings}

Majority of the expensive operations in DNNs involve Convolutions and matrix multiplication operations. These operations can be represented using einsums as follows:

\begin{equation} \label{eqn:gemm}
    O_{m,n}=\Sigma_{k}A_{m,k}\times B_{k,n}
\end{equation}

\vspace{-5.5mm}

\begin{equation} \label{eqn:conv}
    O_{n,h,w,k}=\Sigma_{c,r,s}I_{n,h\text{{+}}r,w\text{{+}}s,c}\times W_{r,s,c,k}
\end{equation}

\autoref{eqn:gemm} represents matrix multiplication with M,K and K,N as the dimensions of the input matrices and M,N as the dimensions of the output matrix.
~\autoref{eqn:conv} represents Convolution layer, with H and W as the feature map height and width respectively, R and S as the filter height and width respectively, N as the batch size and C and K as the number of input and output filters respectively.

These operations can be represented as loop nests as shown below.

%\vspace{-1mm}
\begin{small}
\begin{center}
  \begin{verbatim}
1. for n in range(N):       
2.  for h in range(H):      
3.   for w in range(W):
4.    for k in range(K):
5.     for c in range(C):
6.      for r in range(R):
7.       for s in range(S):
8.        O(n,h,w,k)+=I(n,h+r,w+s,c)*W(r,s,c,k)
\end{verbatim}  
\end{center}
\end{small}

\textit{Dataflow} is used to describe the loop transformations for staging the operations in space and time on a spatial accelerator. Here, the order of the temporal loops in the above example, is NHWKCRS where N is the outermost loop. The dataflow determines the compute utilization and the locality of the tensors in the memory hierarchy, thus choice of the right dataflow is crucial for performance and energy efficiency. In this work, term "dataflow" is to refer to hardware agnostic loop transformations.

One example of a loop transformation is shown below-
%\vspace{-1mm}
\begin{small}
    
  \begin{verbatim}
 1. for n in range(N):       
 2.  for h in range(H):      
 3.   for w in range(W):
 4.    for k in range(K):
 5.     for c1 in range(C/C0):
 6.      for r in range(R):
 7.       for s in range(S):
 8.        parfor c0 in range(C0): 
 9.         c=c1*C0+c0
10.        O(n,h,w,k)+=I(n,h+r,w+s,c)*W(r,s,c,k)
\end{verbatim}  
\end{small}

\vspace{-1mm}

Moreover, C is spatially parallelized across multiple processing elements or PEs. C0 is based on the PE array size. Hardware dependent loop transformations determine the complete \textit{mapping}. We discuss the space for inter-operation dataflows and mappings in~\autoref{sec:dg}

%In~\autoref{sec:dg}, we discuss in detail, the Inter-Operation dataflows.

%In the above example, the dimension C is tiled, as in broken down into tiles of size C0, which is referred to as \textit{Tiling}. Tiling can be spatial or temporal. In the above example, parfor on c0 here represents a spatial tile.A combination of \textit{Dataflow} and \textit{Tile sizes} determines is known as a \textit{Mapping}. \textit{Mapping} is determines the schedule of the workload on the accelerator and directly affects the performance and the energy efficiency of the workload on the accelerator.

\subsection{Inter-Operation Pipelining}
Layer-by-layer computation leads to the whole feature map being written into the memory which leads to high occupancy of the data in the memory hierarchy and also causes excess roundtrip memory accesses to fetch the data for the next layer. To overcome this, prior works have proposed inter-operation pipelining, also known as inter-layer pipelining or layer fusion or inter-operation pipelining where a portion of the feature map is produced and used by the next layer, decreasing the occupancy of the data inside the memory hierarchy and consequently reducing the overall roundtrip memory accesses. Since the next layer consumes only a portion that the previous layer produces, the dataflows of both the layers must be staged to ensure that, thus leading to an interdependence of dataflows~\cite{garg2021understanding}. %We discuss the inter-operation dataflows in more detail in~\autoref{sec:dg} and ~\autoref{sec:violin}.
%\subsection{Inter-operation Pipelining Latency Model}
~\label{sec:model}

\insertWideFigureScaled{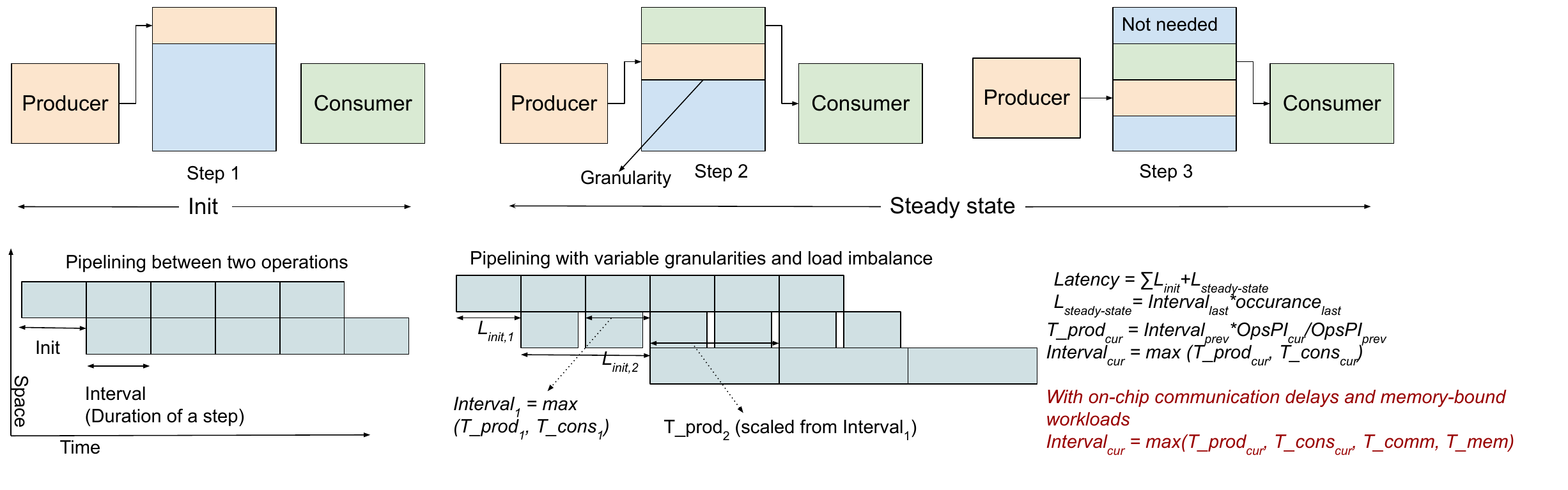}{Inter-operation pipelining between producer and consumer and latency equation for pipelines with arbitrary depth. For fine-grained pipelining, where tiles are small enough to fit inside the local memory of PEs, the data can also be moved through the NoC as opposed to accessing the global buffer.}{0.83}

~\autoref{fig:pipeline-concepts} shows the producer-consumer relationship in inter-operation pipelining. The producer produces a portion of the intermediate data in one timestep which is consumed within the next timestep with producer producing the next piece of data in parallel. Once the data has been consumed, that data is no longer needed.
The first stage where the consumer has not started consuming is referred to as "init" while the rest of the stages are collectively referred to as "steady-state". We refer to the duration of the timestep as "interval". The portion of the intermediate data produced/consumed in a timestep is referred to as "granularity" here. 

As we can see in~\autoref{fig:pipeline-concepts} pipelining can be expressed as a waterfall diagram with vertical axis showing different operations and horizontal axis showing time. With variable granularities and load imbalance, we compute producer side delay as the previous interval delay normalized by the ratio of the number of operations in the current interval and the previous interval. The interval delay is the maximum of the producer side and the consumer side delays. The overall latency is the summation of all the interval delays once (which accounts for all the init delay) and the steady state delay of the last operation. These subtleties are captured in the equations in~\autoref{fig:pipeline-concepts}.

\insertFigure{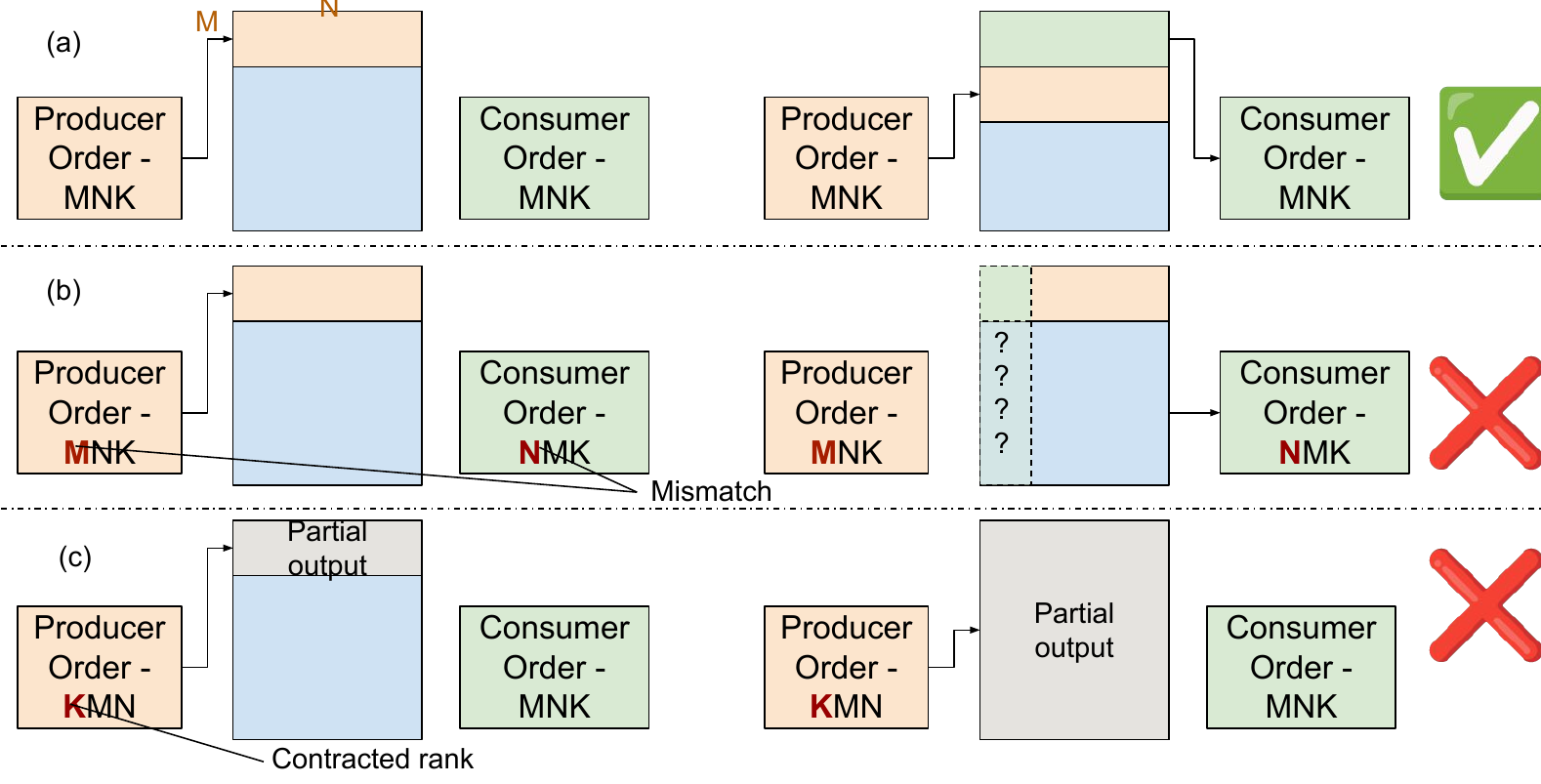}{Conditions for inter-operation pipelining. (a) Conditions met (b) Violation of the same outermost loop (c) Contracted rank of the producer in the outermost loop}

%\RG{Violin C code}

%\subsection{Conditions for Inter-operation Pipelining}
\label{sec:conditions}
Conditions (shown in~\autoref{fig:Conditions}) for making inter-operation pipelining between producer and consumer possible are-
\squishlist
\item For a tensor shared between the producer and the consumer, atleast the outermost loop should be the same. This is needed to divide the producer and consumer into stages. %The granularity of staging is determined by the loop order and is described later in~\autoref{alg:gran}.
\item The contracted rank should not be the outermost rank for the producer, since complete sums are needed earlier for consumption. Similarly, the unshared rank of the consumer must not be in the outermost loop since it would nullify the benefits of pipelining by having to use the complete intermediate tensor in the inner loops multiple times.
\squishend

\vspace{-3mm}

\insertWideFigureScaled{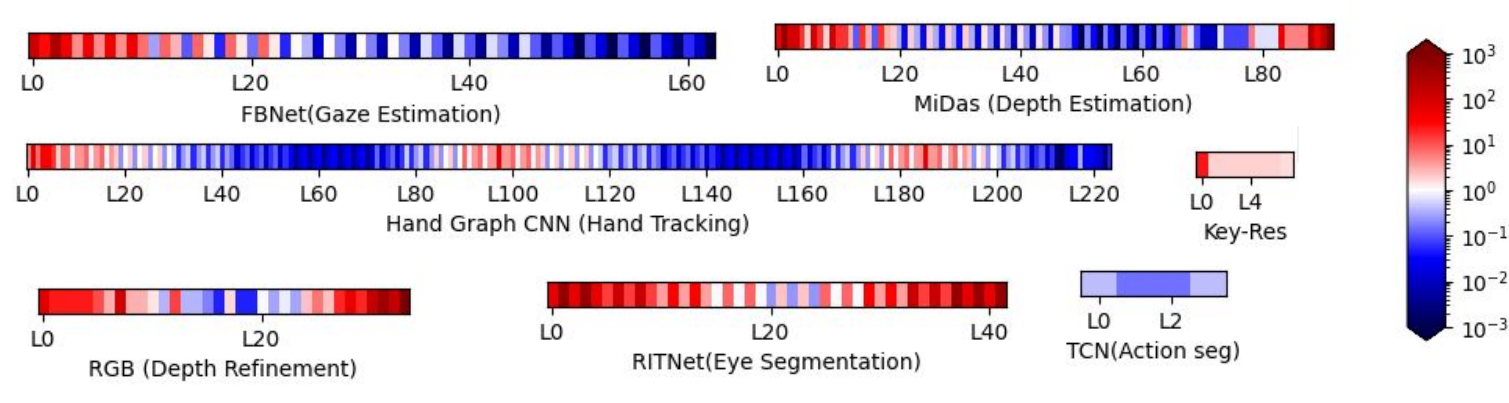}{Activation/weight ratios of CNN tasks inside XRBench~\cite{xrbench}. Layers in red have larger activations, than weights, and the blue ones have larger weights. This excludes the skip connection activation traffic.}{0.9}

\insertWideFigureScaled{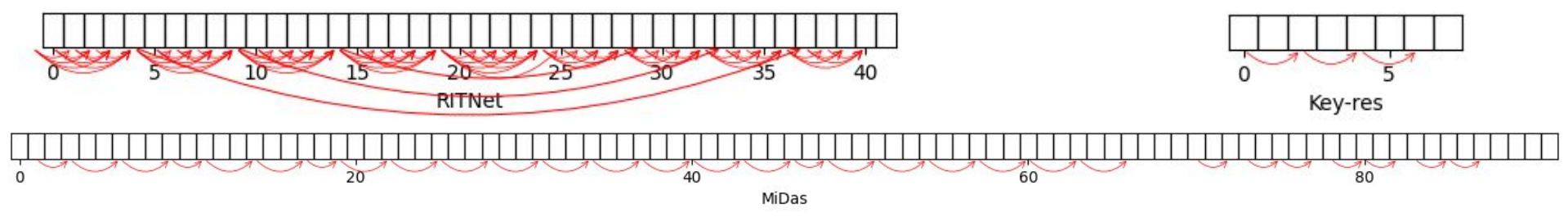}{Skip connections between convolutional layers in XR-bench CNN models}{0.92}

\subsection{Related Work}
\vspace{-1mm}
\label{sec:background}

Various prior works like Atomic Dataflow~\cite{atomic-dataflow}, Stream~\cite{genetic-pipeline}, HyGCN~\cite{yan2020hygcn}, OMEGA~\cite{garg2021understanding}, TANGRAM~\cite{tangram} and FusedCNNs~\cite{fusedcnn} work on inter-operation pipelining between two tensors. Some mapping frameworks like CoSA~\cite{cosa} and GAMMA~\cite{kao2020gamma} focus on layer-by-layer computation but explore multiple dataflows on a flexible accelerator.%~\autoref{tables:related} shows a summary of prior works.

%\vspace{-2mm}
 %\input{tables/related_work}

Most of the prior works on pipelining do not consider both variable pipeline depth and variable granularity together. We show the prior works in terms of their support for fine grained pipelining and depth awareness in~\autoref{tables:related-pipeline}. Please note, that each work presents a different set of contributions, and our work primarily focuses on interconnect and spatial organization. This is the first work, within DNN accelerators that also explores finer-grained spatial organization of layers on PEs (\autoref{fig:orgs}).

%\vspace{-2mm}
\begin{table}[ht!]
\begin{scriptsize}
\begin{center}
\caption{Related work table showing prior works on inter-layer pipelining. Regarding orange checkmark, - 1) TileFlow canonically supports variable depth but fixes it within an application, and 2) HyGCN has two granularities.%\TK{this table still says violin!}
}
\label{tables:related-pipeline}

{%\tt
\begin{tabular}{|c|c|c|c|}\hline
 \textbf{Prior Work} & \textbf{Variable} & \textbf{Variable} & \textbf{Contibution}\\
 & \textbf{Depth} & \textbf{Granularity} &
 \\\hline

 Atomic  & \greencheck & \redcheck & Inter-operation\\
 Dataflow~\cite{atomic-dataflow} & & & mapper
 \\\hline

 Stream~\cite{genetic-pipeline} & \redcheck & \greencheck & Heterogeneous \\
 & & & Dataflow Framework\\\hline

  HyGCN~\cite{yan2020hygcn} & \redcheck & \orangecheck & Accelerator \\\hline

  EnGN~\cite{liang2020engn} & \redcheck & \redcheck & Accelerator\\\hline
   OMEGA~\cite{garg2021understanding} & \redcheck & \greencheck & Cost model  \\\hline

   TANGRAM~\cite{tangram} & \greencheck & \redcheck & New dataflow  \\\hline

    FusedCNN~\cite{fusedcnn} & \redcheck & \redcheck  & Accelerator  \\\hline

    SET~\cite{isca-pip} & \greencheck & \redcheck & Mapper  \\\hline
 
TileFlow~\cite{tileflow} & \orangecheck & \greencheck & Mapper \\\hline

SIMBA~\cite{simba} & \greencheck & \redcheck & Mapper\\\hline

    \DataflowName &  \greencheck & \greencheck & Spatial organization,  \\    
    &&& NoC and Dataflow\\\hline

\end{tabular}
}
\end{center}
\end{scriptsize}
\vspace{-0.1cm}
%'Sp' prefix implies support for zero skipping. IP=Inner Product, OP=Outer Product, Gust=Gustavsons}
\vspace{-0.45cm}
\end{table}

\subsection{Motivation toward Flexible Pipelining: Heterogeneity in AR/VR models}

As XR-bench~\cite{xrbench} shows, DNN models have high heterogeneity in terms of kinds of layers (convolution, depthwise convolution, GEMM, RPN, ROIAlign), in terms of DAG (skip connections with various reuse distances and densities in a block), layer dimensions etc. \autoref{fig:AW} shows the activation/weight ratios of layers within the XR-bench models. These ratios range six orders of magintude from activation-dominant to weight-dominant layers. Moreover, unlike traditional models like ResNet~\cite{resnet}, the location of activation heavy and weight heavy layers inside the model isn't predictable. \autoref{fig:skip} shows the skip connections within five X-R bench models. These skip connections vary in reuse distance and density. For example, RITNet has dense skip connections of multiple reuse distances and midas one skip connection per block with varying reuse distance. Therefore, high heterogeneity in layers requires more attention to finding the right depth and granularity of pipelining, and a flexible NoC that can efficiently execute each scenario.

%\insertFigure{granularity-loop}
%{Impact of loop order on pipeline granularity\vspace{-3mm}}

\section{Pipelining Design-space}
\label{sec:dg}

%\TODO{Add introductory paragraph}

%\subsection{How Mapping Influences Granularity}
%\subsection{Overview of the design-space}

The pipelining dataflow space consists of four aspects - depth, intra-operator dataflow, granularity and spatial organization.

%\TODO{(Definitions)} 

\noindent 
\textbf{Depth. } Refers to the number of layers being pipelined at a time. The model is divided into "pipeline segments" of various depths.

\noindent 
\textbf{Intra-operator Dataflow. } Refers to the dataflow of the individual operator.

\noindent 
\textbf{Granularity. }  Refers to the size/portion of the intermediate tensor consumed by the consumer. For example, granularity of pipelining in~\autoref{fig:pipeline-concepts} is one row.

\noindent 
\textbf{Spatial Organization. } Patterns in which different layers can be arranged on the PEs.~\autoref{fig:orgs} shows examples of different spatial organization strategies.

DNN models can be represented as a DAG of layers. We divide the model into segments of variable depths, which is dependent on shapes of multiple layers and DAG dependencies. We choose the intra-operator dataflow, which depends on the shape of that layer. Based on the intra-operator dataflow, we determine the granularity at which we can pipeline a pair of producer and consumer, within the confines of the depth. Then, we determine the right spatial organization strategy, which depends on the depth and the granularity of pipelining.

The first three namely, depth, granularity and intra-operator dataflow are agnostic of specific hardware details like NoC topology etc. Spatial organization on the other hand, comprises of mapping as it depends upon specific hardware details like NoC topology, PE array dimensions etc.

%Dataflows and tile sizes for a layer can greatly influence pipeline granularity.

\subsection{Factors Affecting Pipelining Depth}
\label{sec:imp-depth}

The choice of pipelining depth impacts inter-operation vs intra-operation reuse. Right pipeline depth depends on two main factors - A/W ratios and skip connections

\textbf{A/W ratios:} Deeper pipeline implies reusing intermediate activations, but at the same time increasing the memory footprint of weights from multiple layers as~\autoref{fig:motivation-1} also shows. Therefore, for pipeline depth of D, weight memory footprint is $\sum_{i=l}^{l+D} W_i$ where $l$ is the first layer of that segment.
Activation memory footprint on the other hand is reduced significantly in comparison, its $A_l+A_{l+D}+\sum_{i=l+1}^{l+D-1} Granularity_i$, where granularity is the portion of the intermediate matrix between the producer and the adjacent consumer. If certain pairs of layers are fine-grained pipelined, the granularity component can be taken care of by PE-to-PE communication leaving the footprint $A_l+A_{l+D}$. Thus larger depth implies, more activations in the middle can simply be skipped from calculation. Although, in case of weights, larger depth implies incurring the footprint of D layers throughout the execution, at all times, reducing the available tile size for weights. \textit{Thus, layers with large activation/weight ratios benefit from deep pipelining while the ones with small ratios benefit from shallow or no pipelining.}

\textbf{Skip connections:} Skip connections were specific to ResNet~\cite{resnet} but have become a norm, and are used in various DNN models~\cite{midas,eyeseg,keyword,fbnet,fbnetv3}. Skip connections can vary in density as well as reuse distance. The activation footprint in presence of the skip connections changes to $A_l+A_{l+D}+\sum A_i$ such that $i\notin(l,l+d)$ and there exists skip connection between $i$ and $j\in(l,l+D)$. This includes both outgoing and incoming skip connections. Thus, for deeper pipelining, the instances of skip connections from outside $(l,l+D)$ are likely to be lower. Therefore, presence of skip connections skews towards deeper pipelining, to absorb those connections within the depth of pipelining. %Outgoing skip connections also favor more pipelining depth since cutting the depth before the destination of the outgoing skip connection ends up increasing the overall memory footprint.

%Another factor that influences the choice of depth is a non-GEMM layer, for example, RCNN has an ROIAlign phase which does not pipeline with the GEMM operation. In this work, however, we only use this factor to reduce the restrict the depth of pipelining. Transformers have activation-activation operations, which we consider in a separate segement from activation-weight operations. So, batched MMs have a pipeline depth of 2.

\subsection{Factors Affecting Intra-operator Dataflow}

The performance of intra-operator dataflow depends primarily on the tensor dimensions~\cite{eyeriss2016isca,kwon2019understanding}, since the purpose is to get reuse. For the off-chip dataflow for operators with extreme A/W ratios, larger tensors should be stationary and the smaller tensors should be streaming, as they can stream from on-chip.

\insertWideFigure{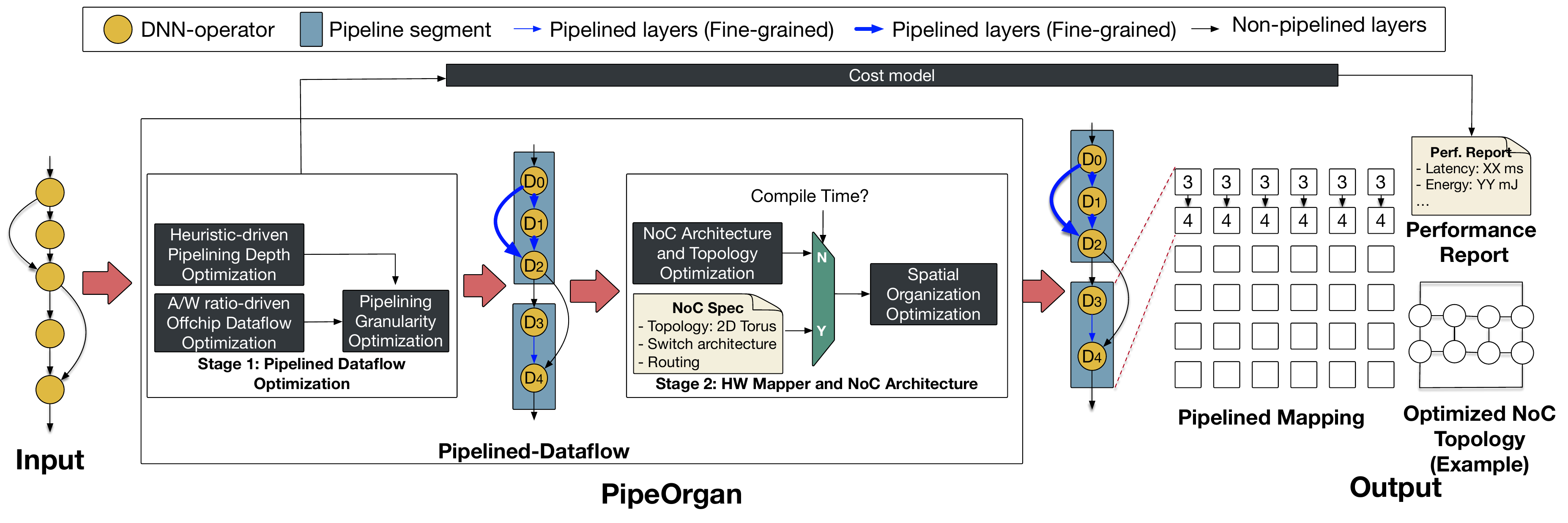}{Entire flow, including mapping heuristics,~\DataflowName and~\NocNamenospace. %\HK{Need to increase the font size. Also, need to reduce white space.}
}

\subsection{Factors affecting Pipelining Granularity}

\label{sec:imp-gran}

Granularity is determined by inta-operator dataflow. For a finer-grained pipelining, the tensor should be consumed in the same order in which it is being produced. For example, in convolutions, the finest grained pipelining is between the pair NHWKCRS-NHWCKRS, given that the data is being consumed exactly as produced. However, the pair NHWKCRS and NHKWCRS has a coarser granulairty since layers can only be staged by NH. 
Similarly for a GEMM, for example, MNK-MKN is the finest grained pipelining possible while MNK-MNK is a coarser grained pipelining.%~\autoref{fig:granularity-loop} shows this visually.
Depending on the loop order, we can calculate the portion of the intermediate tensor and hence the granularity.

Tile sizes can have an impact of granularity particularly if they are unequal. Lets consider the pair NHWKCRS-NHWKCRS. Just based on the loop order, the pipeline granularity is that of a filter. However, if, for example tile size of H is different, the producer and consumer will only be synchronize only when $LCM(Tile\_H_{producer},Tile\_H_{consumer})$ rows have been computed. Hence regardless of the loop order, difference in tile sizes can affect the granularity of pipelining.

\subsection{Factors affecting Spatial Organization}
 The right spatial organization depends on the pipeline depth, which determines how many different layers are allocated at a time, and on pipeline granularity, which determines whether the inter-layer communication should fine-grained or coarse-grained. Prior works divide the PE array into blocked chunks based on the depth and assigned a layer to the chunk, however this is inefficient for fine-granularity pipelining. Therefore we propose a new class of mappings where the layers can be organized flexibly on the PE array.

%~\autoref{fig:orgs} shows examples of spatial organization strategies for pipeline depth of 4. 
In~\autoref{sec:pipeorgan}, we focus on spatial organization in more detail.

\insertWideFigure{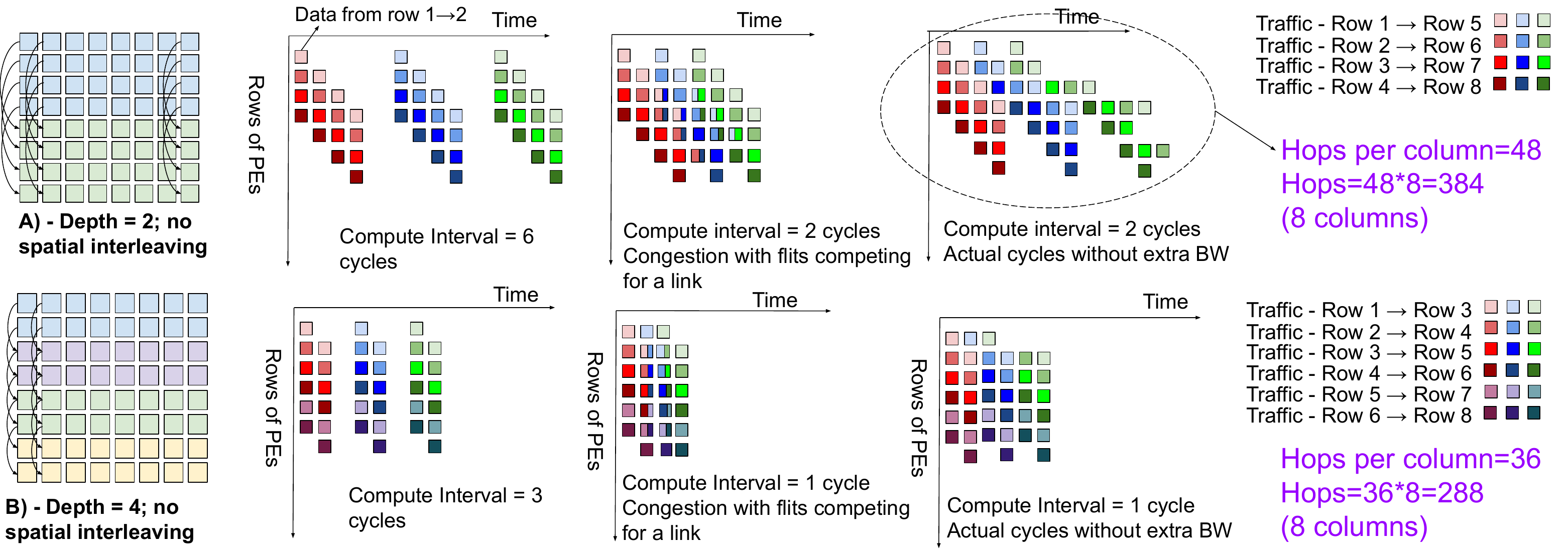}{Fine-grained inter-operation pipelining with coarse-grained (blocked) spatial allocation for depth=2 and depth=4.}

\section{\DataflowNamenospace: Flexible Spatial Organization}
\label{sec:pipeorgan}
\vspace{-1mm}

~\autoref{fig:flow} shows the whole flow of~\DataflowNamenospace. The first stage involves pipelined dataflow optimization, and the second stage involves hardware-aware mapping and NoC architecture.

\textbf{Stage 1: Pipelined Dataflow Optimization} This stage involves using heuristics to choose the right depth and intra-operator dataflow, which then helps determine granularity of pipelining.

\textbf{Stage 2: HW mapping and NoC architecture}: In the second stage on the depth and granularity, we determine the spatial organization strategy to determine the complete mapping. At design-time we study the traffic patterns of various pipelined-dataflows on various spatial organization strategies, identify bottlenecks in the mesh and propose~\NocNamenospace. At compile-time, we determine the spatial organization strategy based for~\NocName based on pipelined-dataflow. We discuss the details of~\NocName in~\autoref{sec:noc}.

\subsection{Pipelined Dataflow}
\label{sec:stage1}
First stage of~\DataflowName takes in an input DAG and uses heuristics to (a) partition the whole model into segements of flexible depth, (b) determine intra-operation dataflows and (c) 

\textbf{Determining Depth}: In this work, we determine depth of a segment (starting at layer $l$) by comparing the memory footprints $A_l+A_{l+D}$ with $\sum_{i=l}^{l+D} W_i$ (see~\autoref{sec:imp-depth}), increasing the value of D. We stop adding more depth, the moment $\sum_{i=l}^{l+D} W_i$ is greater. In case of skip connections, we also add additional activations due to skip connections, thus skip connections skew the descision towards deeper pipeline. We also cut the depth if we encounter a complex layer like ROIAlign. The depth is also limited by the size of the substrate. The maximum depth we consider is $\sqrt{numPEs}$. 

\textbf{Determining Intra-operation Dataflows (Loop order)}: 
Intra-operation dataflows greatly influence pipelining and even the ability to pipeline as~\autoref{sec:imp-gran} shows. Ideal intra-operation dataflows can depend on layer shapes as prior works~\cite{eyeriss2016isca,kwon2019understanding} have shown. For the scope of the paper, we simply choose a few dataflows depending on the ratio of the weight and activation volumes. In case of larger weights, we use weight stationary dataflow, where ranks from weights form the outermost loop, to get more reuse on weights. This dataflow is not friendly to pipelining. While for the activation-heavy layers, we choose the activation stationary dataflow. Depending on how large activation is compared to the weight, we decide weather to make the dataflow completely activation stationary (for example, NHWKCRS) or we allow some reuse on weights (for example, NHKCWRS). We validate our heuristic on XR-bench usage scenarios. We are able to achieve the best possible arithmetic intensity\footnote{Best-case arithmetic intensity is obtained by considering only cold misses.} in case of 99.94\% of the layers with on-chip buffer size of 512KB and 97.2\% of the layers with on-chip buffer size of 256KB. Note that this only determines the order of dimensions in the outer loops, not including spatial parallelism and spatial tiling.

\textbf{Determining Finest Possible Granularity Based on Loop Order}
We determine granularity from intra-operation dataflows using~\autoref{alg:gran} model in this section.
\vspace{-1mm}
\begin{algorithm}[ht!]
%\label{algo:inter-op}
\SetInd{0em}{1em}
\caption{Determination of granularity from intra-operation dataflows (within a pipeline segment)}\label{alg:gran}
\begin{footnotesize}
Granularity = Output size\\
pipnest=0\\ 
\textcolor{gray}{\%Within each pipeline segment}\\
    \For{layer $\in$ CUR to CUR+DEPTH-1} 
{   
\For{loop $\in$ 0 to NUM\_UNCONTRACTED\_LOOPS}
{
\If{(loop==0 || tilesz[layer][loop-1]==tilesz[layer+1][loop-1])\\
\&\& (looppair(layer,layer+1)==(N,N)||(H,H)||(W,W)||(K,C))}
{
$Granularity=Granularity/Dimension[layer][loop]*lcm(tilesz[layer][loop],tilesz[layer+1][loop])$\\
$pipnest+=1$
}
}
}

\end{footnotesize}

\end{algorithm}
%\vspace{-1mm}

The algorithm compares the loop pairs to determine if they can be fused and does the same till the loop nests are fusible. However it stops if there is a mismatch in the tile size as discussed in~\autoref{sec:imp-gran}. This step determines the finest possible granularity, but the granularity can also change depending on the parallelization strategy determined in the spatial organization.

%Once we have the depth, granularity and intra-operation dataflows, we determine the best possible spatial organization for each scenario. We do it across a myriad of traffic patterns, caused by skip connections of varying reuse distances and varying densities, unequal allocation of PEs to different layers, layer shapes etc.

\insertWideFigure{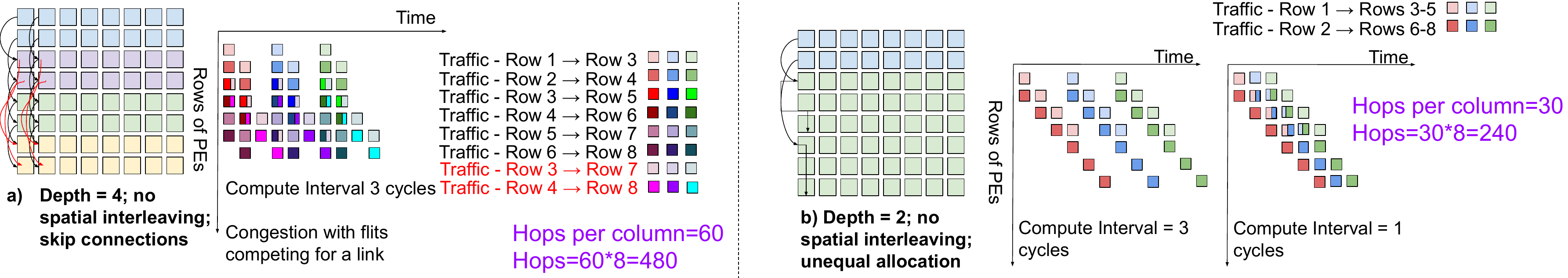}{Effect of (a) skip connection (b) unequal PE allocation due to load balancing on traffic pattern}
%\insertFigure{skip_connections}{Effect of skip connections on traffic.}

%\insertFigure{load_balance}{Effect of load imbalance in inter-operation traffic patterns.}

\insertWideFigure{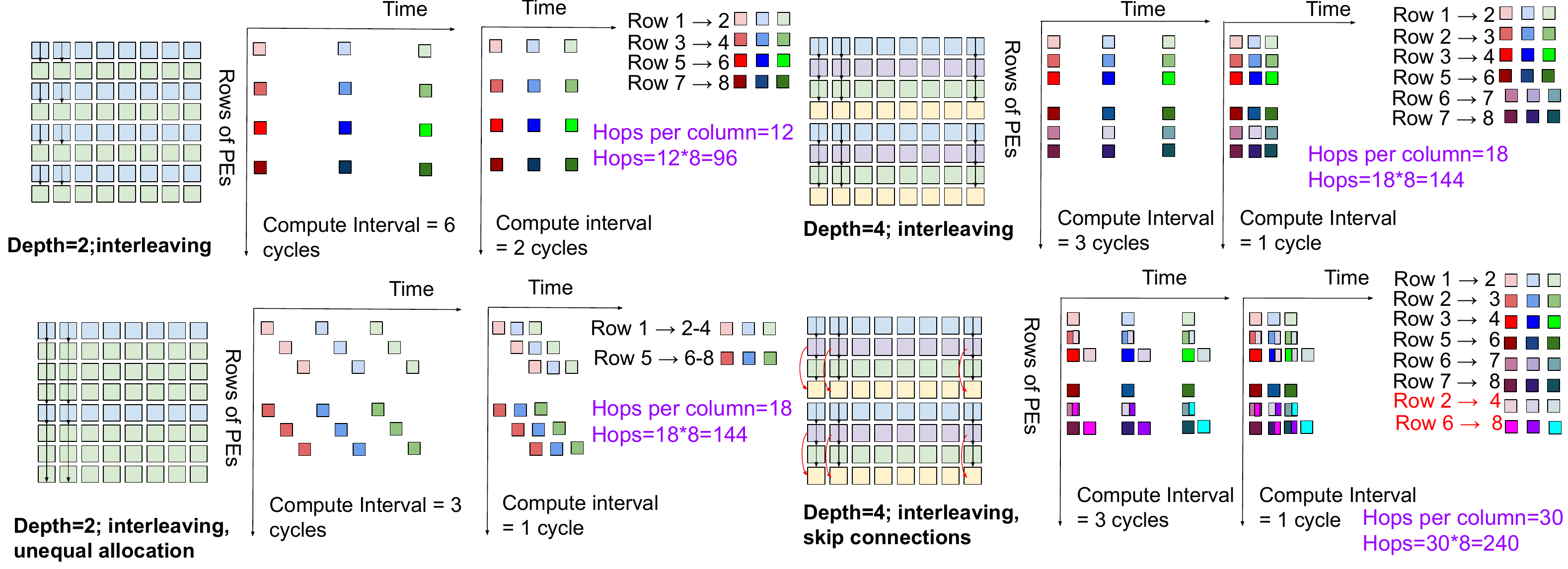}{Congestion-free traffic with 1-D interleaving (fine-striped configuration). Significant hop reduction is observed corresponding to counterparts from~\autoref{fig:fine_grained_patterns} to~\ref{fig:skip2}.}

\insertWideFigure{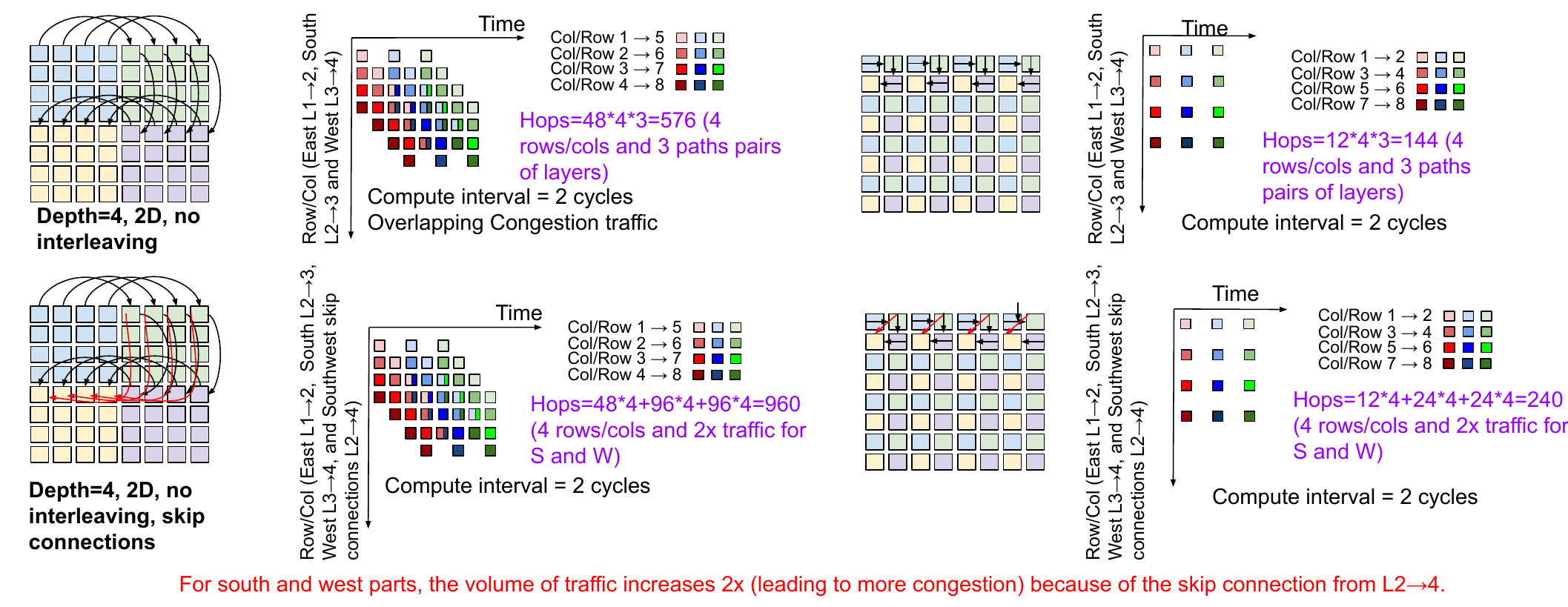}{2-D spatial organization with and without interleaving and with and without skip connections. Pipelining with 2-D allocation can be broken down into multiple 1-D paths. In this example, eastern path spans communication from layer 1 to 2, southern path  - layer 2 to layer 3 and western path - layer 3 to layer 4. Skip connection requires traversal along multiple directions.%\HK{Need to increase the font size; make it consistent with Figs 9 and 10. You can reduce some white space in this figure to get a larger font size.}
}

\subsection{HW Mapping: Determining Spatial Organization Strategy}
\label{sec:stage2}

 To determine spatial organization, we use the depth to ascertain the number of layers to organize spatially. We allocate number of PEs for each layer based on the ratio of MACs.
 
 We determine the arrangement of those layers based on granularity. We compare the total register file (RF) size with the granularity of pipelining.
If $RF_{total
} < Granularity$, the data is moved through the Global Buffer. In mappings with coarse granularity of pipelining, the intermediate data is moved through the Global Buffer (GB). This is always done in a blocked organization. 

On the contary if the total register file size of the producer is larger than the granularity, the spatial organization is decided based on how fine the granularity is, relative to the PE register file. For example, the finest possible granularity can be executed on a checkerboard organization or via sequential pipelining. If the granularity is almost at par with total producer register file size, then pipelining can be done in a blocked organization. The number of PEs involved on the producer side is determined by $Granularity/RF\_per\_PE$. 

The key idea is that once the granularity is ascertained, we allow each pipeline interval (defined in~\autoref{fig:pipeline-concepts}) to be mapped flexibly to allow for optimal intra-operation reuse. Fine-grained spatial organization for coarse-grained pipelining constrains the parallelization and tiling strategies that the individual layers use. For example, in case of sequential pipeline fine-grained checkerboard allocation and sequential pipelining, only the part of the intermediate tensor that is produced, would be consumed. For example, if the activations parallelize K and W dimension, they also have to be consumed in the same way, with fine-grained spatial allocation.

Hence coarser-grained pipelining also should use coarser-grained spatial organization. Additionally, 1-D vs 2-D organization is decided based on the depth and on the reuse within the pipeline stage (whether its more along one dimension, or along multiple dimensions). Once the spatial organization strategy is decided, PEs could be allocated to the layers in ratios that ensure load balancing and maximum utilization. Parallelization strategy is also decided at this step, which could potentially increase the granularity from stage 1, but this change does impact the spatial organization decision.

\subsection{HW Mapping: Design-time Traffic Analysis}

\autoref{fig:fine_grained_patterns} shows the traffic patterns generated by fine grained pipelining with blocked 1D spatial organization. Here, we compare the total hop time against the interval time. The compute interval time stems from the temporal reduction that needs to be done within the PEs to produce an output that can be consumed. If this time is greater than the hop count, congestion does not happen. However, if this time is less, it leads to congestion. This is due to the contention caused by the new traffic getting generated at a faster rate. On resolving this congestion, we find that the latency is limited by the hop count rather than the compute interval. This traffic pattern also has a high overall hop count. Moreover, coarse-grained pipelined dataflows benefit from intra-operation reuse. %Within DNN workloads, the effect of congestion is much more pronounced in case of cloud scale, and cases of congestion are still possible on edge. However a biggest concern for the edge is the energy spent due to large hop counts in blocked spatial organization (upto $num\_rows/2$).

~\autoref{fig:skip2}a) shows additional congestion that is caused by skip connections in residual blocks of ResNet. This would be even more detrimental, in case of RITNet where there is a skip connection to each later layer inside the block, with highly activation-heavy layers.

\autoref{fig:skip2}b) shows, a case where PEs for the producer and consumer are unequally allocated. The traffic hotspot is at the boundary of the producer and the consumer, so it is hard to pinpoint the exact row of PEs where it happens. A common example of this is ResNet, with 1x1 and 3x3 filter sizes and varying number of output channels inside a residual block.

Finally, \autoref{fig:1Dinterleaving} shows, how 1-D interleaving can avoid congestion by co-locating the producer and the consumer tiles, which is beneficial for fine-grained exchange between the layers. It also shows the traffic benefits in case of skip connections and unequal allocation. Significant reduction in congestion and hop count is observed compared to the blocked organization counterparts in~\autoref{fig:fine_grained_patterns}-\ref{fig:skip2}. 

~\autoref{fig:2Dinterleaving} compares blocked and fine-grained spatial organization for 2-D allocation, with depth=4. In 2-D allocation, the traffic can be broken down into multiple sets of paths involving 1-D communication. In this example, communication from layer 1 to 2 is along east in the top half, and four rows are involved. Communication from layer 2 to 3 is along the south in the right half and four columns are involved. similarly communication from layer 3 to 4 is along west in the bottom half and four rows are involved.
Skip connections involve traversal along both southern and western paths doubling the traffic for those.

Notice that coarse grained pipelining incurs less congestion even with blocked organization, however, it incurs high hop count. Reducing the hop count would require us to organize the coarse grained pipelined layers in a tightly coupled manner, which reduces intra-op mapping flexibility and efficiency. This motivates us to propose~\NocName to account for the traffic patterns, in order to not trade-off intra-operation flexibility for on-chip energy. We summarize the bottlenecks caused by various scenarios on mesh in~\autoref{table:noc}.

\begin{scriptsize}
\begin{table}[h]
\begin{scriptsize}
    
    \centering
      \caption{Summary of mesh bottlenecks observed in~\autoref{fig:fine_grained_patterns}-\ref{fig:2Dinterleaving}.}
    \label{table:noc}
    \begin{tabular}{|c|c|c|}
            \hline

        \textbf{Cause} & \textbf{Effect} & \textbf{Prevalent in} \\\hline
       Many long overlapping paths  & High Congestion & Blocked 1D and 2D \\\hline
       Many long overlapping paths  & High hop energy & Blocked 1D and 2D\\\hline
       Extra BW for skip connections & High congestion & All organizations \\\hline
        Extra hops with skip connections & High hop energy & All configurations \\\hline
        Routing in multiple directions & Higher hop energy & 2D organizations\\\hline
    \end{tabular}
  \end{scriptsize}
\end{table}
\end{scriptsize}

%\RG{List of different kinds of communication patterns}

\insertFigure{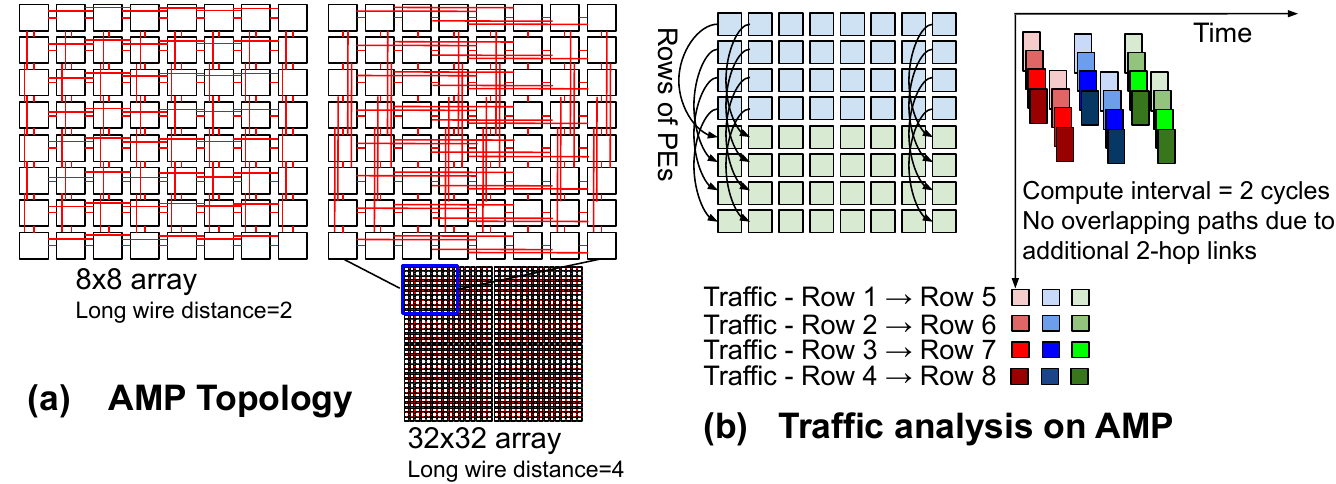}{(a)\NocName topology (b) Traffic analysis on depth=2, no spatial interleaving with~\NocName topology.}

\subsection{\NocName Topology}
\label{sec:noc}
%\textbf{Bottlenecks in Mesh:}
~\autoref{table:noc} shows the bottlenecks with the mesh topology. These bottlenecks are caused by overlapping paths that could go all the way through a row/column in case of coarse-grained spatial organization (as~\autoref{sec:pipeorgan} discusses, fine-grained spatial organization can resolve this, but is not always possible). This bottleneck is made worse by skip connections, as they introduce extra congestion. Flattened butterfly~\cite{flat} adds links to increase the bandwidth but is an overkill and the number of links increase in complexity from $O(N)$ to $O(NlogN)$. SMART~\cite{smart} adds single-cycle multi-hop support to mesh, and can help in reducing energy, however, it does not resolve congestion by increasing the available bandwidth. 

\begin{comment}
%\textbf{Topology:}
Key characteristics of traffic patterns to consider here are-

\noindent
(1) Because of variable depth, there is no specific hotspot, therefore links need to be added uniformly.

\noindent
(2) Because of skip connections, larger depth does not necessarily minimize long paths, which necessitates direct communication along long paths.

\noindent
(3) Congestion and hops in a mesh increase with longer paths as they span multiple links, thus long path links are beneficial.

\noindent
(4) Moreover, longer path links reduce the hop count and also avoid congestion since there are direct long distance paths without traffic contending for the links, so essentially the channel bandwidth requirement itself is reduced.

\noindent
(5) Too long links would not help with congestion concentrated in PEs not covered by the long link bypassing them.
%To this end, we propose a new NoC that uses the topology that keeps the number of links to $O(N)$. We also discuss the corresponding routing and the microarchitectural modifications.

\end{comment}

%\textbf{Topology}
~\autoref{fig:amp}a shows the~\NocName topology for $8\times 8$ array.
We modify the conventional mesh topology by adding wires of length $Round(\sqrt{\frac{Rows}{2}})$\footnote{We choose this as its the geomean of single hop and $\frac{Rows}{2}$ hop case in~\autoref{fig:fine_grained_patterns}.}, in each PE, from each direction. This limits the length of the wires and it scales by $O(\sqrt{N})$ (the wire length spans 4 PEs for a 32$\times$32 PE array and 8 PEs for a 64$\times$64 PE array). By the virtue of having long enough wires itself, the congestion decreases because of less overlapping paths, without links going all the way unlike torus.~\NocName increases the number of links compared to mesh by under 2x.
This also reduces the hop count.~\autoref{fig:amp}b shows the resulting reduction in congestion and hop count.

%\textbf{Routing:} As~\autoref{fig:fine_grained_patterns}-\ref{fig:2Dinterleaving} show, we use one direction routing in case of 1-D allocation, while we use zigzag routing(also used in TANGRAM~\cite{tangram}) for coarse-grained 2D allocation. For fine-grained allocation, we the same allocation within localized zones of multiple layers. By doing this, we avoid any chance of deadlock. For the new links, while we keep the direction same, we make a choice between the mesh link and the long link, based on whether the distance in terms of PEs is greater than or less than $Round(\sqrt{\frac{Rows}{2}})$ PEs. For example, in a $32\times32$ array, the longer wires would be 4 PEs apart. The path from row 1 to row 17 would be 1$\rightarrow$5$\rightarrow$9$\rightarrow$17, while the path from row 1 to row 19 would be 1$\rightarrow$5$\rightarrow$9$\rightarrow$17$\rightarrow$18$\rightarrow$19.

%\textbf{Switch architecture:} \NocName topolgy does not require major switch architecture change except for a mux with a control signal indicating the link choice based on path ahead.

%\RG{Topology and microarchitecture figure}

%\subsection{NoC Configuration}

%\RG{Detailed figure with communication patterns, demonstrating the choice of spatial organization.}
%\input{05-Arch}
\section{Experimental Methodology}

\label{sec:expt}

%We describe our experimental methodology for demonstrating the efficiency of \DataflowName across baseline dataflows. We focus on the inter-cluster communication and the DRAM accesses which is same regardless of microarchitectural parameters.

\subsection{Evaluation Framework}
\label{sec:simulation}

\insertWideFigure{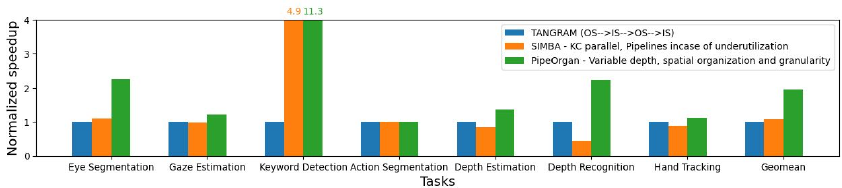}{End-to-end performance benefits for each task in XR-bench~\cite{xrbench}. Results are normalized to TANGRAM-like dataflow. Higher is better.%\HK{Let's add geomean bars at the right end.}
%\RG{Depth estimation left}
}

\insertWideFigure{dram}{End-to-end normalized DRAM accesses for each task in XR-bench~\cite{xrbench}. Results are normalized to TANGRAM-like dataflow. Lower is better. %\HK{Let's add geomean bars at the right end.}
%\RG{Depth estimation left}
}

%\insertWideFigureScaled{energy}{Overall energy benefits for each usage scenario in XR-bench~\cite{xrbench}.\RG{placeholder}}{0.85}

We develop an in-house simulation framework for evaluating the spatial organizations.

\textbf{\DataflowNamenospace:} The framework models \DataflowName stage 1 using the heuristics described in detail in~\autoref{sec:stage1}. Based on the pipeline depth ie, number of different layers, and the pipeline granularity, the framework determines the right spatial organization strategy at compile time as~\autoref{sec:stage2} explains, and evaluate performance and energy. At design-time, the framework evaluates the performance and energy of different spatial organization strategies for different stage 1 outputs, for traffic analysis and NoC design. 

\textbf{Performance modeling:} The framework models the communication energy and latency cost using in-house NoC simulator to model traffic patterns, topology and routing to compute the hops and estimate the congestion. The NoC simulation automates the NoC and traffic analysis visually shown in~\autoref{fig:fine_grained_patterns}-\ref{fig:2Dinterleaving}. We obtain the producer and consumer compute interval latency based on~\autoref{fig:pipeline-concepts}. We also factor in the additional stalls  due to limited on-chip memory and limited memory bandwidth. Using compute, communication and memory components, we are able to obtain inter-layer pipelining, based on~\autoref{fig:pipeline-concepts}.

\subsection{Workloads and Datasets}

We primarily evaluate \DataflowName and~\NocName on XR-bench~\cite{xrbench} CNN tasks, capturing different tasks related to eye tracking, hand tracking, keyword spotting, world locking, object detection, action segmentation etc. As~\autoref{fig:AW}-\ref{fig:skip} show, these workloads have wide variery of layer shapes and different DAG structures due to skip connections. %These also consist of a variery of kernels including CONV, DWCONV, 1D CONV, pooling etc. % (Table 2 of XR-bench paper~\cite{xrbench}).
%We evaluate on all the usage scenarios discussed in XR-bench which consist of a instances of different tasks.
%(Table 1 of XR-bench paper~\cite{xrbench}).
%For example, the VR gaming scenario consists of hand tracking and eye tracking tasks, and AR assistant scenario consists of keyword detection, speech recognition, semantic and object segmentation, and world locking tasks.

%We also observe the specific benefits of~\DataflowName and~\NocName at larger scale where congestion is more prominent and likely for wrong spatial organizations. We use MLPerf-cloud~\cite{mlperf} for large scale evaluation.

%\input{tables/workload.tex}

\label{sec:dataset}
%We evaluate our methodology and the \DataflowName mapping strategy for Conjugate Gradient and GNN (GNN results in~\autoref{sec:GNN}). We obtain the sparse matrices of the Conjugate Gradient Datasets from Suitesparse~\cite{suitesparse} for scientific problems like structural problems, stencils, acoustics etc. We obtain GNN graphs from OMEGA~\cite{omega}.

\subsection{Baselines}
\label{sec:baseline}

We compare the end-to-end performance of~\DataflowName on~\NocName against the dataflows used by TANGRAM~\cite{tangram} and SIMBA~\cite{simba}, for each XR-bench task~\cite{xrbench}. TANGRAM-like dataflow alternates between output stationary and input stationary, thus uses fine-grained pipelining with depth=2. SIMBA like dataflow parallelizes input and output channels and does pipelining only when these two dimensions cannot utilize the substrate.% We also do specific comparsions between mesh and~\NocName and~\DataflowName and blocked organization. %We also use another baseline with flexible depth but no fine-grained spatial organization to focus on the benefits of~\DataflowNamenospace. We call it "Variable depth".
%We also demonstrate the energy benefit of spatial organization strategy exclusively by comparing~\DataflowName fine-grained spatial organization against TANGRAM's dataflow~\cite{tangram}, and the benefit of~\NocName on coarse-grained . (\autoref{fig:pipestuff}). We also show the benefits of~\NocName's topology and architecture.

We also show the depth and granularity as an output of stage 1 for each of the XR-bench~\cite{xrbench} task in~\autoref{fig:depth} and~\ref{fig:gran}.
%\input{tables/dataflow_config.tex}

%We compare multiple dataflow configurations described in~\autoref{table:dataflow_config}. SEQ-Flex dataflow has the lowest possible DRAM accesses for individual layers but it considers the output tensor of each operation ending up in the DRAM and the input coming from the DRAM. SEQ-Overflow dataflow considers that the data is written in the SRAM and the portion that does not fit is sent to the DRAM. To make the baseline strong and fair, we writeback the tensor in FIFO order (but do not consider reuse distance based replacement) in SEQ-Overflow. GOGETA-df uses inter-op patterns to reduce pressure on memory and finds the individual loop order with pipelining and also avoids swizzle\_penalty. GOGETA-map also uses the scalable tiling strategy that reduces inter-cluster communication overhead in addition to GOGETA-df. GOGETA-df and GOGETA-map also include the tensor allocation strategy of FIFO order write and reuse distance based replacement in~\autoref{sec:tornado}. Ideal, represents the DRAM accesses with perfect reuse.

\subsection{Architecture Parameters}
\label{sec:params}

We consider the architecture parameters shown in~\autoref{table:config}.

\begin{scriptsize}
    \begin{table}[h!]
\begin{scriptsize}
    
  \begin{center}
  \caption{Configurations for workloads and architecture} 
  %\Rav{There is space for remarks too if needed}}
  %\TK{@Raveesh - by x do you mean you can do either s or t for that dimension? So does that mean each row is actually multiple datapoints? Thats confusing.}\Rav{It means that the datapoint that we choose for evaluation of that dataflow can have varaible tile sizes for dimension marked by X. S and T on the other hand mean that its NECESSARILY spatial or temporal}}
  \label{table:config}
  \begin{centering}
  \begin{tabular}{|l|l|}
    \hline
    \textbf{Parameter} & \textbf{Value}  \\
    \hline
    %Seq & $V\times F$ & $t_{AGG}+t_{CMB}$\\
    %\hline
    Bytes per word/element & 1B (8 bits)\\\hline
    PE array size & 32$\times$32\\ 
    \hline
    PE dot product size\footnote{No reduction for grouped convolutions} & 8\\\hline
  %  PE array size at scale (\autoref{fig:scale}) & 256$\times$256\\ \hline
  % PE dot product size at scale (\autoref{fig:scale}) & 32\\\hline
    SRAM capacity & 1MB \\\hline
 %  SRAM capacity at scale (\autoref{fig:scale}) & 8MB\\ \hline
%    Off-chip bandwidth & \\\hline
 %   On-chip bandwidth & \\\hline
  %  Inter-cluster bandwidth & \\\hline
   Memory bandwidth & 256 GB/s\\\hline
 %   Memory bandwidth at scale (\autoref{fig:scale}) & 1TB/s\\\hline
    %Total L2 capacity & 32MB\\ \hline
    %Total L3 capacity & 256MB\\ \hline
  \end{tabular}
\vspace{-3mm}
\end{centering}
\end{center}
\end{scriptsize}

\end{table}
\end{scriptsize}

%\input{tables/config.tex}
%\insertWideFigure{DRAM}{DRAM data movement in Megabytes for N (x-axis) = 1, 8, 16 and SRAM sizes, 1MB, 4MB and 16MB. We cut the Y-axis at a lesser value for certain plots due to large disparity between SEQ-Flex and GOGETA-map.} %\RG{Only SEQ-Flex is real, rest are placeholders for now.}}

%~\autoref{table:config} describes the architecture configuration we use for the evaluation and the parameters we sweep. We run workloads of different sizes and nnz's and we also sweep the $N$ rank that corresponds to number of simultaneous initial guesses and the SRAM size to capture different scenarios with varying ratios of tensor and SRAM sizes.
%We consider a clustered architecture, each cluster of 1024 PEs backed by its own SRAM slice.

\insertWideFigure{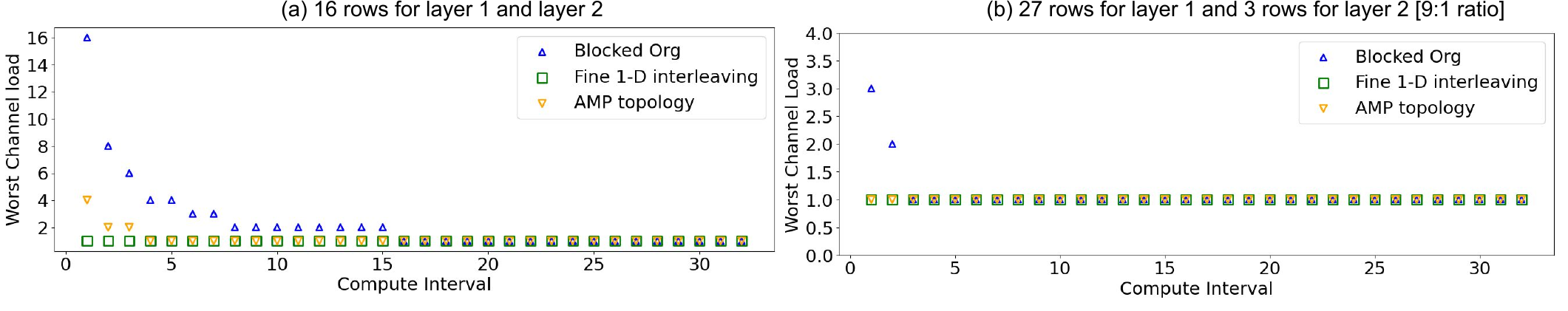}{Worst case channel load as a function of compute interval for 1-D spatial allocation with depth=2 for $32\times32$ mesh/\NocName. We compare blocked organization, \DataflowName fine 1-D organization and \NocNamenospace.}

\insertWideFigure{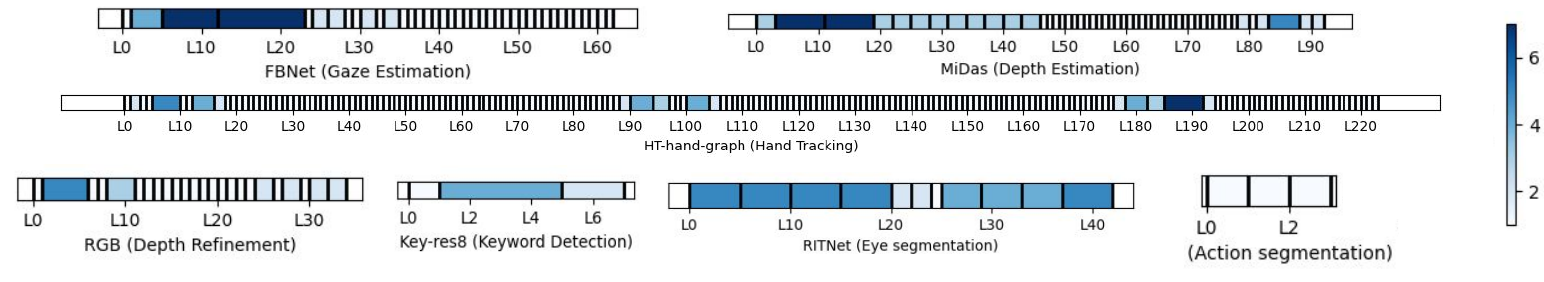}{Depths of XR-bench CNN tasks. L0 stands for Layer 0 and so on.}

\insertWideFigure{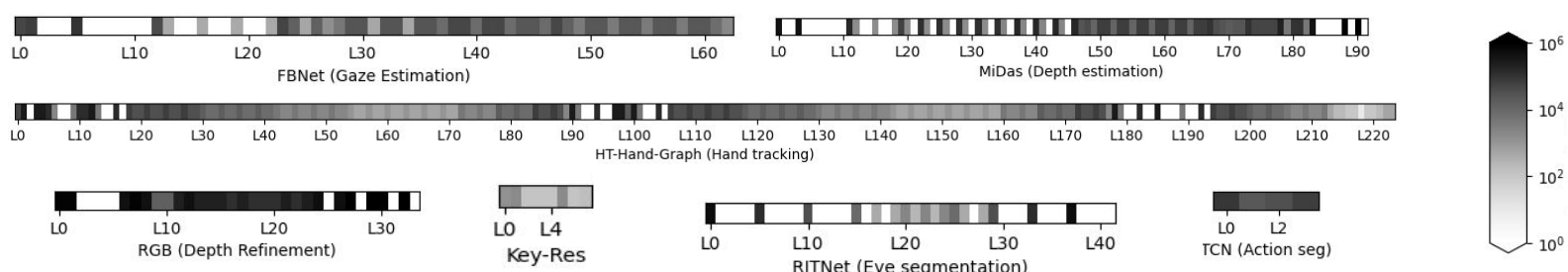}{Finest possible granularity based on depth and intra-operation dataflows across CNN tasks
. L0 stands for Layer 0 and so on.}

\section{Results}
\label{sec:eval}

%\subsection{Validation results}
%\RG{Hard coded RTLs}

%\subsection{\DataflowName v/s Prior Dataflows}
%\subsubsection{AR/VR scenarios}
%\subsubsection{Large Language Models}

\subsection{Performance}
\autoref{fig:perf} shows the end-to-end performance benefits of~\DataflowName over SIMBA-like~\cite{simba} and TANGRAM-like~\cite{tangram} dataflows for each task. SIMBA-like dataflow pipelines if one layer is unable to utilize the substrate. SIMBA experiences latency in cases where parallelizing input channels and output channels is not sufficient. Other major source of latency is load imbalance, specially in cases with different filter sizes. TANGRAM-like dataflow uses fine-grained pipelining alternating between output stationary and input stationary. However, blocked spatial allocation leads to congestion, hence deteriorating the performance, as it becomes on-chip NoC bound. KD-resnet in particular, is the most affected in case of TANGRAM-like, since the compute interval duration is 1-cycle, and blocked organization is too coarse.~\DataflowName uses fine-grained spatial organization to avoid this issue. This is similar to the behavior observed in~\autoref{fig:fine_grained_patterns}. Action segmentation and hand tracking are mostly weight heavy, with large channels, and therefore do not favor pipelining. Gaze estimation and depth estimation do better with deeper pipelining in the activation heavy regions, given that DWCONV layers are memory bound.~\DataflowName exploits reuse through flexible depth and spatial organization. All in all,~\DataflowName gains geomean 1.95x in performance (this includes all the layers and not just the layers that benefit from pipelining).
\subsection{Normalized DRAM accesses}
\vspace{-2mm}

\autoref{fig:dram} shows end-to-end normalized DRAM accesses, which are normalized to TANGRAM-like. High DRAM access reduction was achieved on eye segmentation due to flexible depth which absorbs the dense skip connections. Similarly, gaze estimation and depth estimation have memory-bound layers, where~\DataflowName is able to reduce memory accesses in earlier segments. All in all, DRAM accesses were reduced by 31\% geomean. Note that this is end-to-end reduction, which also includes non-pipeline friendly layers (which in some cases, might also have larger tensors, and hence more accesses).

\subsection{Congestion analysis:~\DataflowName and~\NocName}
%\RG{Placeholder}
%\TODO{\TODO{\TODO{}}}

~\autoref{fig:congestion} shows the variation of congestion with compute interval and spatial organization strategies. We show 1-D allocation with depth=2 since we observe most of the congestion in those cases, specially in TANGRAM-like dataflow. Two highly observed cases were equal allocation, and unequal allocation with $1\times1$ anf $3\times3$ filters. Blocked organization has a large delay per packet for equal allocation case. The overall interval delay is $Worst\,case\, channel\,load \times Compute\,Interval$. For compute interval of 2 cycles, the overall communication delay increases by a factor of 8, as a result its 16. Fine-grained 1-D interleaving resolves this by avoiding congestion since it brings the consumer closer. Likewise,~\NocName topology, by virtue of long links, reduces the congestion delay and only incurs congestion when compute interval is below 4 cycles. The unequal allocation case incurs lower latency compared to the equal allocation counterpart, nevertheless, blocked organization is still more likely to cause congestion.

\subsection{Pipeline Depth and Granularity}

\subsubsection{Pipelining Depth across Tasks}
\autoref{fig:depth} shows the depths of CNN layers obtained after applying the depth heuristic, for the entire models. %We observe variable depths in various tasks.
Eye segmentation~\cite{eyeseg} task has the most regions with deep pipelining, primarily because of the high A/W ratios and skip connections. Keyword detection~\cite{keyword} in particular prefers pipelining despite nominal A/W ratios because of skip connections. Depth estimation~\cite{midas} also has multiple deep pipelined regions. This is primarily because of DWCONV layers, which have a high A/W ratio, given that, weights are only along one channel. Moreover, these layers actually need pipelining since these layers are memory bound compared to a CONV with same number of multiplications. %Similarly, gaze estimation also has DWCONV layers, and the early few layers have deep pipelining. 

\subsubsection{Finest Pipelining Granularity across Tasks}~\autoref{fig:gran} obtain the finest possible granularities assigned by stage 1, for the entire CNN models. However, this does not include the final granularities after the spatial organization is finalized. Granularity also depends on activation size, thus some cases appear fine-grained despite no pipelining. High-depth regions can have long regions of finer granularity.

%\subsection{Benefits of~\DataflowName at Scale}

%\subsubsection{Scaling inputs and wieights}

%\vspace{-1mm}
\section{Conclusion}
\label{sec:discussion}

Modern DNN application domains consist of multiple models with vastly different layer shapes, layer dependencies and types of operations used. The right depth and granularity of inter-operation pipelining or lack thereof depends heavily on these model characteristics. Prior works on pipelining miss out on the opportunity to take advantage of fine-grained inter-operation pipelining opportunities and regions are allocated to layers in a course-grained inter-layer manner, which has a significant on-chip communication overhead. In this work, we propose~\DataflowNamenospace, a new class of spatial organization strategies for inter-operation pipelining, and a systematic methodology to choose the right depth, granularity and consequently the spatial organization. We also propose~\NocNamenospace, a topology that reduces the hops and congestion for coarse-grained spatial allocation.

\section*{Acknowledgement}

Part of this work was supported by ACE, one of the seven centers in JUMP 2.0, a
Semiconductor Research Corporation (SRC) program sponsored by DARPA.
\clearpage
\bibliographystyle{IEEETranS} %%%%% 
%\bibliographystyle{plain}
\bibliography{refs}
%\clearpage
%\input{AppendixA.tex}

%\input{Appendix}

\end{document}